\documentstyle[12pt,epsfig,epsf]{article}

\catcode`\@=11
\long\def\@makefntext#1{ 
\protect\noindent \hbox to 3.2pt {\hskip-.9pt
$^{{\ninerm\@thefnmark}}$\hfil}#1\hfill} 

 \def\@makefnmark{\hbox to 0pt{$^{\@thefnmark}$\hss}}  

\def\ps@myheadings{\let\@mkboth\@gobbletwo
\def\@oddhead{\hbox{} 
\rightmark\hfil\ninerm\thepage}
\def\@oddfoot{}\def\@evenhead{\ninerm\thepage\hfil 
\leftmark\hbox{}}\def\@evenfoot{}
\def\sectionmark##1{}\def\subsectionmark##1{}}
\def\AJ{{\it Astrophys.J.} }
\def\AJL{{\it Ap.J.Lett.} }
\def\AJS{{\it Ap.J.Supp.} }
\def\AM{{\it Ann.Math.} }
\def\AP{{\it Ann.Phys.} }
\def\APJ{{\it Ap.J.} }
\def\APP{{\it Acta Phys.Pol.} }
\def\ASAS{{\it Astron. and Astrophys.} }
\def\BAMS{{\it Bull.Am.Math.Soc.} }
\def\CMJ{{\it Czech.Math.J.} }
\def\CMP{{\it Commun.Math.Phys.} }
\def\FP{{\it Fortschr.Physik} }
\def\HPA{{\it Helv.Phys.Acta} }
\def\IJMP{{\it Int.J.Mod.Phys.} }
\def\JMM{{\it J.Math.Mech.} }
\def\JP{{\it J.Phys.} }
\def\JCP{{\it J.Chem.Phys.} }
\def\LNC{{\it Lett. Nuovo Cimento} }
\def\SNC{{\it Suppl. Nuovo Cimento} }
\def\MPL{{\it Mod.Phys.Lett.} }
\def\NAT{{\it Nature} }
\def\NC{{\it Nuovo Cimento} }
\def\NP{{\it Nucl.Phys.} }
\def\PL{{\it Phys.Lett.} }
\def\PR{{\it Phys.Rev.} }
\def\PRL{{\it Phys.Rev.Lett.} }
\def\PRTS{{\it Physics Reports} }
\def\PS{{\it Physica Scripta} }
\def\PTP{{\it Progr.Theor.Phys.} }
\def\RMPA{{\it Rev.Math.Pure Appl.} }
\def\RNC{{\it Rivista del Nuovo Cimento} }
\def\SJPN{{\it Soviet J.Part.Nucl.} }
\def\SP{{\it Soviet.Phys.} }
\def\TMF{{\it Teor.Mat.Fiz.} }
\def\TMP{{\it Theor.Math.Phys.} }
\def\YF{{\it Yadernaya Fizika} }
\def\ZETF{{\it Zh.Eksp.Teor.Fiz.} }
\def\ZP{{\it Z.Phys.} }
\def\ZMP{{\it Z.Math.Phys.} }

\newcounter{sectionc}\newcounter{subsectionc}\newcounter{subsubsectionc}
\renewcommand{\section}[1] {\vspace{0.6cm}\addtocounter{sectionc}{1}
\setcounter{subsectionc}{0}\setcounter{subsubsectionc}{0}\noindent
        {\bf\thesectionc. #1}\par\vspace{0.4cm}}
\renewcommand{\subsection}[1] {\vspace{0.6cm}\addtocounter{subsectionc}{1}
        \setcounter{subsubsectionc}{0}\noindent
        {\it\thesectionc.\thesubsectionc. #1}\par\vspace{0.4cm}}
\renewcommand{\subsubsection}[1] {\vspace{0.6cm}\addtocounter{subsubsectionc}{1}
        \noindent {\rm\thesectionc.\thesubsectionc.\thesubsubsectionc.
        #1}\par\vspace{0.4cm}}

\newcounter{appendixc}
\newcounter{subappendixc}[appendixc]
\newcounter{subsubappendixc}[subappendixc]

\renewcommand{\appendix}[1] {\vspace{0.6cm}
        \refstepcounter{appendixc}
        \setcounter{figure}{0}
        \setcounter{table}{0}
        \setcounter{equation}{0}
        \renewcommand{\thefigure}{\Alph{appendixc}.\arabic{figure}}
        \renewcommand{\thetable}{\Alph{appendixc}.\arabic{table}}
        \renewcommand{\theappendixc}{\Alph{appendixc}}
        \renewcommand{\theequation}{\Alph{appendixc}.\arabic{equation}}
        \noindent{\bf Appendix \theappendixc #1}\par\vspace{0.4cm}}

\def\abstracts#1{{
        \centering{\begin{minipage}{30pc}\tenrm\baselineskip=12pt\noindent
        \centerline{\tenrm ABSTRACT}\vspace{0.3cm}
        \parindent=0pt #1
        \end{minipage} }\par}}

\renewenvironment{thebibliography}[1]
        {\begin{list}{\arabic{enumi}.}
        {\usecounter{enumi}\setlength{\parsep}{0pt}
\setlength{\leftmargin 1.25cm}{\rightmargin 0pt}
         \setlength{\itemsep}{0pt} \settowidth
        {\labelwidth}{#1.}\sloppy}}{\end{list}}

\newcounter{itemlistc}
\newcounter{romanlistc}
\newcounter{alphlistc}
\newcounter{arabiclistc}

\newcommand{\fcaption}[1]{
        \refstepcounter{figure}
        \setbox\@tempboxa = \hbox{\tenrm Fig.~\thefigure. #1}
        \ifdim \wd\@tempboxa > 6in
           {\begin{center}
        \parbox{6in}{\tenrm\baselineskip=12pt Fig.~\thefigure. #1 }
            \end{center}}
        \else
             {\begin{center}
             {\tenrm Fig.~\thefigure. #1}
              \end{center}}
        \fi}

\newcommand{\tcaption}[1]{
        \refstepcounter{table}
        \setbox\@tempboxa = \hbox{\tenrm Table~\thetable. #1}
        \ifdim \wd\@tempboxa > 6in
           {\begin{center}
        \parbox{6in}{\tenrm\baselineskip=12pt Table~\thetable. #1 }
            \end{center}}
        \else
             {\begin{center}
             {\tenrm Table~\thetable. #1}
              \end{center}}
        \fi}

\def\@citex[#1]#2{\if@filesw\immediate\write\@auxout
        {\string\citation{#2}}\fi
\def\@citea{}\@cite{\@for\@citeb:=#2\do
        {\@citea\def\@citea{,}\@ifundefined
        {b@\@citeb}{{\bf ?}\@warning
        {Citation `\@citeb' on page \thepage \space undefined}}
        {\csname b@\@citeb\endcsname}}}{#1}}

\newif\if@cghi
\def\cite{\@cghitrue\@ifnextchar [{\@tempswatrue
        \@citex}{\@tempswafalse\@citex[]}}
\def\citelow{\@cghifalse\@ifnextchar [{\@tempswatrue
        \@citex}{\@tempswafalse\@citex[]}}
\def\@cite#1#2{{$\null^{#1}$\if@tempswa\typeout
        {IJCGA warning: optional citation argument
        ignored: `#2'} \fi}}

\def\fnt#1#2{\footnotetext{\kern-.3em
        {$^{\mbox{\sevenrm #1}}$}{#2}}}

 1
 1
 1

\font\tenbf=cmbx10
\font\tenrm=cmr10
\font\tenit=cmti10

\font\ninerm=cmr9

\begin{document}
\begin{flushright}
CERN-TH/98-43\\
hep-th/9802057
\end{flushright}
\vspace{0.4in}

\centerline{\tenbf A SIMPLE/SHORT INTRODUCTION TO PRE-BIG-BANG
PHYSICS/COSMOLOGY}
\baselineskip=22pt
\baselineskip=16pt
\vspace{0.8cm}
\centerline{\tenrm GABRIELE VENEZIANO}
\centerline{\tenit Theory Division, CERN,}
\baselineskip=12pt
\centerline{\tenit 1211 Geneva 23, Switzerland}
\baselineskip=13pt
\vspace{0.9cm}
\abstracts{A  simple, non-technical introduction to the pre-big bang scenario is given,
emphasizing physical motivations,  considerations, and
consequences over formalism.}
\def\AJ{{\it Astrophys.J.} }
\def\AJL{{\it Ap.J.Lett.} }
\def\AJS{{\it Ap.J.Supp.} }
\def\AM{{\it Ann.Math.} }
\def\AP{{\it Ann.Phys.} }
\def\APJ{{\it Ap.J.} }
\def\APP{{\it Acta Phys.Pol.} }
\def\ASAS{{\it Astron. and Astrophys.} }
\def\BAMS{{\it Bull.Am.Math.Soc.} }
\def\CMJ{{\it Czech.Math.J.} }
\def\CMP{{\it Commun.Math.Phys.} }
\def\FP{{\it Fortschr.Physik} }
\def\HPA{{\it Helv.Phys.Acta} }
\def\IJMP{{\it Int.J.Mod.Phys.} }
\def\JMM{{\it J.Math.Mech.} }
\def\JP{{\it J.Phys.} }
\def\JCP{{\it J.Chem.Phys.} }
\def\LNC{{\it Lett. Nuovo Cimento} }
\def\SNC{{\it Suppl. Nuovo Cimento} }
\def\MPL{{\it Mod.Phys.Lett.} }
\def\NAT{{\it Nature} }
\def\NC{{\it Nuovo Cimento} }
\def\NP{{\it Nucl.Phys.} }
\def\PL{{\it Phys.Lett.} }
\def\PR{{\it Phys.Rev.} }
\def\PRL{{\it Phys.Rev.Lett.} }
\def\PRTS{{\it Physics Reports} }
\def\PS{{\it Physica Scripta} }
\def\PTP{{\it Progr.Theor.Phys.} }
\def\RMPA{{\it Rev.Math.Pure Appl.} }
\def\RNC{{\it Rivista del Nuovo Cimento} }
\def\SJPN{{\it Soviet J.Part.Nucl.} }
\def\SP{{\it Soviet.Phys.} }
\def\TMF{{\it Teor.Mat.Fiz.} }
\def\TMP{{\it Theor.Math.Phys.} }
\def\YF{{\it Yadernaya Fizika} }
\def\ZETF{{\it Zh.Eksp.Teor.Fiz.} }
\def\ZP{{\it Z.Phys.} }
\def\ZMP{{\it Z.Math.Phys.} }
\def\laq{\ \raise 0.4ex\hbox{$<$}\kern -0.8em\lower 0.62
ex\hbox{$\sim$}\ }
\def\gaq{\ \raise 0.4ex\hbox{$>$}\kern -0.7em\lower 0.62
ex\hbox{$\sim$}\ }
\def\half{\hbox{\magstep{-1}$\frac{1}{2}$}}
\def\quarter{\hbox{\magstep{-1}$\frac{1}{4}$}}
\def\NPB{{\em Nucl. Phys.} B}
\def\PLB{{\em Phys. Lett.}  B}
\def\PRL{{\em Phys. Rev. Lett. }}
\def\PRD{{\em Phys. Rev.} D}
\def\MPL{{\em Mod. Phys. Lett.}  A}

\def\gappeq{\mathrel{\rlap {\raise.5ex\hbox{$>$}} {\lower.5ex\hbox{$\sim$}}}}

\def\lappeq{\mathrel{\rlap{\raise.5ex\hbox{$<$}} {\lower.5ex\hbox{$\sim$}}}}

\def\beq{\begin{equation}}
\def\eeq{\end{equation}}
\def\bea{\begin{eqnarray}}
\def\eea{\end{eqnarray}}

\parskip 0.3cm\parskip 0.3cm



\section{Introduction}

It is commonly believed (see e.g.\cite{Hawking}) that the Universe
-- and time itself -- started some 15 billion years ago from some kind
of primordial explosion, the famous Big Bang.  Indeed, the
experimental observations of the red-shift and of the Cosmic Microwave
Background (CMB) lead us quite unequivocally to the conclusion that,
as we trace back our history, we encounter epochs of increasingly high
temperature, energy density, and curvature. However, as we arrive
close to the singularity, our classical equations are known to break
down. The earliest time we can think about classically is certainly
larger than the so-called Planck time, $t_P = \sqrt{G_N \hbar} \sim
10^{-43} \rm{s}$ ($c=1$ throughout).  Hence, the honest answer to the
question: Did the Universe and time have a beginning? is: We do not
know, since the answer lies in the unexplored domain of quantum
gravity.

Besides the initial singularity problem -- and in spite of its
successes -- the hot big bang model also has considerable
phenomenological problems.  Amusingly, these too can be traced back to
the nature of the very early state of the Universe.  Let us briefly
recall why.

General Relativity, together with the cosmological principle (i.e. the
assumption of a homogeneous, isotropic Universe over large scales),
allows us to describe the geometry of space-time through the
Friedmann-Robertson-Walker (FRW) metric (see e.g.\cite{Weinberg}):
\begin{equation}
ds^2 = -dt^2 + a(t)^2~ \left[  {dr^2 \over 1- k r^2} + r^2 d\Omega^2  \right] \; ,
\; d\Omega^2 = d \theta^2 + \rm{sin}^2 \theta d \varphi^2 \; ,
\label{FRW}
\end{equation}
where, as usual, the discrete variable $k=0,1,-1$ distinguishes the
cases of a flat, closed, or open Universe, respectively.  In the
presence of some matter (fluid), described by an energy density $\rho$
and pressure density $p$, the evolution of the Universe is controlled
by the Einstein-Friedmann equations
\begin{eqnarray}
H^2 \equiv \left({\dot{a} \over a} \right)^2 &=& { 8 \pi G \over 3} \rho - { k \over a^2}
\nonumber \\
\dot{H} + H^2 = {\ddot{a} \over a}  &=& - { 4 \pi G \over 3} (\rho + 3 p) \; ,
\label{EF}
\end{eqnarray}
which, together, imply the energy conservation equation:
\begin{equation}
\dot{\rho} = -3 \left({\dot{a} \over a} \right) ( \rho + p) \; .
\label{Econs}
\end{equation}
Given a model for the sources, or, more specifically, given a relation
between $\rho$ and $p$ (so-called equation of state), one can easily
solve these equations and find the usual FRW cosmological solutions.
For the matter- and radiation-dominated cases, respectively,
\begin{eqnarray}
p = 0  &\Rightarrow&  a(t) \sim t^{2/3} , \; \rho \sim a^{-3} \nonumber \\
p = \rho/3  &\Rightarrow&  a(t) \sim t^{1/2} , \; \rho \sim a^{-4}
\label{MRD}
\end{eqnarray}
where, for simplicity, the spatial-curvature term ${ k \over a^2}$ was
neglected.

The so-called horizon problem arises as follows.  The observable part
of our Universe, our present horizon, is given by the distance that
light has travelled since the big bang, or about $10^{28}~ \rm{cm}$.
At earlier times, the horizon was much smaller.  For a hypothetical
observer looking at the sky a few Planck times after the Big Bang, the
horizon was not much bigger that a few Planck lengths, say about
$10^{-32}~ \rm{cm}$.  Instead, as easily seen from (\ref{MRD}), the
portion of space that corresponds to our present horizon was, for that
same hypothetical observer, some $30$ orders of magnitude larger than
the Planck length, or about $1~ \rm{mm}$.  In other words, at that
time, what has nowadays become our observable Universe consisted of
$(10^{30})^3 = 10^{90}$ Planckian-size, causally disconnected regions.
There is no reason to expect that conditions in all those regions were
initially the same, since there was never any thermal contact between
them. Yet, today, all those $10^{90}$ regions make up our observable
Universe and all appear to resemble one another (to within one part in
$10^5$).

Who prepared such a smooth and very unlikely initial state?  Perhaps
God, who picked up a very special point in the huge space of all
possible initial configurations (see, in this connection, a nice
picture in Roger Penrose's book\cite{Penrose}).  If, instead, we do
not accept God's fine-tuning, or, in more scientific terms, we do not
want to attribute homogeneity to some unknown Planckian physics, two
logical possibilities are left to our choice:
\begin{itemize}
\item Time did have its beginning at the big bang, when initial
  conditions were rather random, but a period of superluminal
  expansion (inflation) brought all those $10^{90}$ patches in causal
  contact sometimes between the big bang and the present time.  This
  is the standard {\it post-big bang} inflation paradigm (see
  e.g.\cite{KT}).
\item Time did {\it not} have its beginning with the big bang and some
  pre-big bang physics cooked up a ``good" big bang from a more
  generic (less fine-tuned) initial state.  This is the attitude one
  takes in the so-called { \it pre-big bang} scenario.
\end{itemize}

One can similarly argue that there are two ways of solving the second
major problem of standard cosmology, the so-called flatness problem.
Today, space is, to a very good approximation, Euclidean. If it does
have any spatial curvature (represented by the ${ k \over a^2}$ term
in the cosmological equations), this is of $O(H^2)$, i.e. extremely
small.  Given the solutions (\ref{MRD}) it is easy to check that, in
order to have such conditions today, one has to start, at the Planck
time, with an extremely flat Universe, i.e. with a curvature radius
$a$ some $30$ orders of magnitude larger that the characteristic
length scale at the time, $H^{-1} \sim l_P$.  Again, two possibilities
come to mind: either the Universe was not particularly flat at the
beginning -- and subsequent
inflation stretched out spatial curvature --
or some pre-big bang physics prepared a nice spatially-flat big bang.

Conventional inflation again chooses the first alternative. To
succeed, it needs a weakly-coupled scalar field, the so-called
inflaton, which, some time after the big bang, finds itself {\it
  homogeneously} displaced from the minimum of its potential, and
slowly rolls towards it.  While doing so, if certain conditions are
met, the effective equation of state is $p \sim -\rho$ and the
Universe expands quasi-exponentially. One needs this period of
exponential growth to last for a long enough time for all our
accessible Universe to come in causal contact. This can be achieved at
the price of fine-tuning certain masses or couplings.

My main reservations towards this solution (see\cite{Penrose1} for a
different criticism) are that no one has a convincing model for what
the inflaton ought to be and, even more seriously, that it is not easy
to justify the initial conditions that can provide a sufficiently long
inflationary phase.  One is back somehow to the starting point, since
the conditions at the onset of inflation have to come from a previous
phase, and this inevitably leads us to giving initial conditions in
the mysterious Planckian era. Quantum cosmology has then been invoked;
however, I am also sceptical about present quantum cosmology
arguments\cite{Hawking1} ``predicting" inflation since they are based
on the so-called minisuperspace truncation of the Wheeler--DeWitt
equation.  Such an approach is only justified for a fairly homogeneous
initial universe, which is just what we do {\it not} wish to assume.

How come string theory prefers the second way out?  In order to
explain this I have to open a parenthesis and tell you about some
striking properties of quantum (super)strings.

\section {Three properties of (super)strings}

I will concentrate on just three properties of strings which are
crucial to the understanding of their possible cosmological
implications:
\begin{itemize}
\item{1.} There is a fundamental length scale in string theory,
  providing a characteristic size for strings and thus an ultraviolet
  cutoff. Thanks to this property, superstring theory can be taken
  seriously as a candidate finite quantum theory of gravity (and of
  the other interactions as well).  The fundamental length scale
  $\lambda_s$ is given in terms of the string tension $T$ by the
  formula
\begin{equation}
\lambda_s = \sqrt{\hbar/T}~.
\end{equation}

Actually $\lambda_s$, rather than $T$, is the fundamental parameter of
the theory, providing a meaning for what short and large distances
mean.  When fields vary little over a string length $\lambda_s$ one
recovers a field-theoretic description given by a local Lagrangian
with the smallest number of derivatives.
\item{2.}  Couplings are not God-given constants; they are VEVs which,
  hopefully, become dynamically determined.  In particular, a scalar
  field, the so-called dilaton $\phi$, controls all sorts of
  couplings, gravitational and gauge alike.  Since, in our
  normalizations,
\begin{equation}
T \cdot G_N = l_P^2 /\lambda_s^2  \sim \alpha_{gauge} \sim e^{\phi} \; ,
\label{VEV}
\end{equation}
we see that the weak coupling region is $\phi \ll -1$. By contrast, at
present, $e^{\phi} \sim 1/25$, implying $\lambda_s \sim 10 l_P \sim
10^{-32}~ \rm{cm}$.  In the weak coupling region, perturbative
superstring theory is an adequate description of physics, implying
that the dilaton itself behaves like a massless particle. As such, the
dilaton can easily evolve cosmologically while it is deeply inside the
perturbative region. Precision tests of the equivalence principle
imply, however, that the dilaton has a mass, i.e. that near its
present value $\phi= O(-1)$, its potential has a minimum with finite
curvature.

If both the coupling and derivatives are small, physics is adequately
described by the tree-level low-energy effective action of string
theory, which reads:
\begin{eqnarray}
\Gamma_{eff} &=& \frac{1}{2} \int d^4x \sqrt{-g}~ e^{-\phi}
\left[\lambda^{-2}_s ({\cal R}
+ \partial_\mu \phi \partial^\mu\phi + H^2_{\mu\nu\rho}) +
F^2_{\mu\nu}
+{\rm higher~derivatives} \right]  \nonumber \\&
+& \left[ {\rm higher~orders~in}~e^\phi \right]~,
\label{31}
\end{eqnarray}
where we have included the contributions of the Kalb-Ramond
antisymmetric tensor field through its field strength
$H_{\mu\nu\rho}$. Note the two kinds of corrections alluded to in
(\ref{31}). They intervene, respectively, whenever space-time
derivatives (i.e. energies) or the string coupling $e^{\phi}$ become
appreciable. Equation (\ref{31}) will be our starting point to
describe pre-big bang cosmology.
\item{3.}  Cosmological field equations exhibit new stringy symmetries
  whose most interesting representative (scale-factor duality or SFD)
  acts as follows:
\begin{equation}
a(t) \rightarrow a^{-1}(-t) \;\; ,~ \phi(t) \rightarrow \phi(-t) 
- 2 d ~ \log a(-t) \;\;,
d=3 \; .
\end{equation}
The interest of this duality transformation lies in the fact that it
maps ordinary FRW cosmologies with $H>0$, $\dot{H} < 0$, and
$\dot{\phi} =0$ at $t>0$ into inflationary cosmologies with $H>0$,
$\dot{H} > 0$, and $\dot{\phi} > 0$ at $t<0$.  Actually, the dual
cosmologies are of the so-called super-inflation (or pole-inflation)
type, i.e. have a growing -- rather than a constant
-- Hubble parameter.
Since many of the distinctive consequences of PBB cosmology originate
from this peculiar feature, let me explain in simplified terms where
it comes from  Eqs. (\ref{EF}) imply that, for an expanding
Universe, $\rho$ and $H^2$, being proportional (take for simplicity
the case of $k=0$), decrease together in time. This is true if $G_N$
is constant. In string theory, where $G_N$ is controlled by the
dilaton through Eq. (\ref{VEV}), it is perfectly possible to have a
growing $H$ while $\rho$ decreases, provided $\phi$ is also growing.
\end{itemize}

The suggestion from string theory now becomes almost a compelling one:
Can one put together a standard cosmology at $t>0$ and a dual
cosmology at $t<0$ to generate a single scenario containing
dilaton-driven pre-big bang inflation and FRW post-big bang behaviour?
Since, in such a construct, the Hubble parameter grows for $t<0$ and
decreases for $t>0$, it should reach its maximum at $t=0$, instant
therefore identified with the occurrence of the Big Bang of standard
cosmology. The problem is that this maximum is actually infinite if
one works in the context of the low-energy effective theory, i.e. if
all corrections in (\ref{31}) are neglected.  The pre-(post-) big bang
phases have singularities in the future(past), probably a consequence
of the validity, in that approximation, of the assumption leading to
the Hawking-Penrose singularity theorems.\cite{HP}  However, that
approximation breaks down as soon as the Hubble parameter reaches
values $O(\lambda_s^{-1})$ leading us to expect that the maximal value
of $H$, reached at $t=0$, should be actually of the order of the
fundamental length of string theory.

It is easy, actually, to write down mathematical expressions that
interpolate smoothly between the inflationary and the FRW branches.
For instance, the ansatz
\begin{eqnarray}
\hat{a}(t) &=& \left({t + \sqrt{t^2 +
\lambda_s^2} \over \lambda_s }\right)^{1/2} \nonumber \\
\hat{\phi}(t) &=&  \phi_0 + {3 \over 2}~
\log \left(1 + { t \over \sqrt{t^2 + \lambda_s^2}}\right)
\end{eqnarray}
is easily checked to approach a standard radiation-dominated cosmology
with constant dilaton at $t \gg \lambda_s$ and to a dual
dilaton-driven inflationary cosmology at $t \ll -\lambda_s$. The
question is: Does anything like this come from the true (i.e.
high-curvature and/or loop-corrected) field equations? Leaving this
hard question to the final section, we turn instead to the formulation
of the basic pre-big bang postulate.

\section{The pre-big bang postulate}

Clearly, if we want to use a dual cosmology for the prehistory of the
Universe, given the positive signs of $\dot{H}$ and $\dot{\phi}$, we
have to start from (very?) small initial values for $H$ and
$e^{\phi}$. Although not strictly necessary, we will also impose, for
the sake of simplicity, an almost empty initial Universe. This leads
us to the following basic postulate of PBB cosmology:

The Universe started its evolution from the most simple initial state
conceivable in string theory, its perturbative vacuum. This
corresponds to an (almost)

\vspace{0.2cm}
\begin{centerline}
  {\bf EMPTY, COLD, FLAT, FREE}
\end{centerline}
\begin{flushleft}
  Universe as opposed to the standard
\end{flushleft}
\vspace{0.2cm}
\begin{centerline}
  {\bf DENSE, HOT, HIGHLY-CURVED}
\end{centerline}
\begin{flushleft}
  initial state of conventional cosmology.
\end{flushleft}

For this assumption to make sense I will have to argue that the new
initial conditions are able to provide, at later times, a hot big bang
with the desired characteristics thanks to a long pre-big bang
inflationary phase.  This looks a priori a very hard task, but I will
explain below how it can possibly happen. Before discussing this let
me illustrate, in two figures, the qualitative differences between the
standard (non-inflationary) model, the standard inflationary scenario,
and ours.

In Fig. 1 I am plotting, against cosmic time, the behaviour of the
Hubble parameter $H$ measured in Planck units $M_P \sim 10^{19}~
\rm{GeV}$. Note that, while in the first model $H$ is a concave
function of time, in the second it has a long flat plateau where
$H/M_P \leq 10^{-5}$ (this constraint comes from COBE's data, see
e.g.\cite{KT}).  Finally, in the proposed scenario, $H/M_P$ grows all
the way to a maximal value $O(10^{-1})$, but quickly becomes very
small at both large positive and large negative $t$.

In Fig. 2 we can see the consequences of this behaviour of $H$ on the
kinematics of horizon crossing. During inflation, increasingly small
scales are pushed out of the horizon by the accelerated expansion of
the Universe. However, while in standard inflation (Fig.~2a) larger
scales cross the horizon at slightly larger values of $H$, in the
pre-big bang scenario (Fig.~2b) it is the other way around: the larger
the scale, the smaller the value of $H$ at horizon crossing.  As we
shall see in Sec. 4, the value of $H/M_P$ at horizon crossing is the
determining quantity for evaluating the present magnitude of quantum
fluctuations at different length scales.  Hence the above kinematics
of horizon crossing will have an important bearing on the spectrum of
quantum fluctuations.

Before turning to that, we should discuss how dilaton-driven inflation
sets in during the pre-big bang phase.

\hglue 2.0cm \epsfig{figure=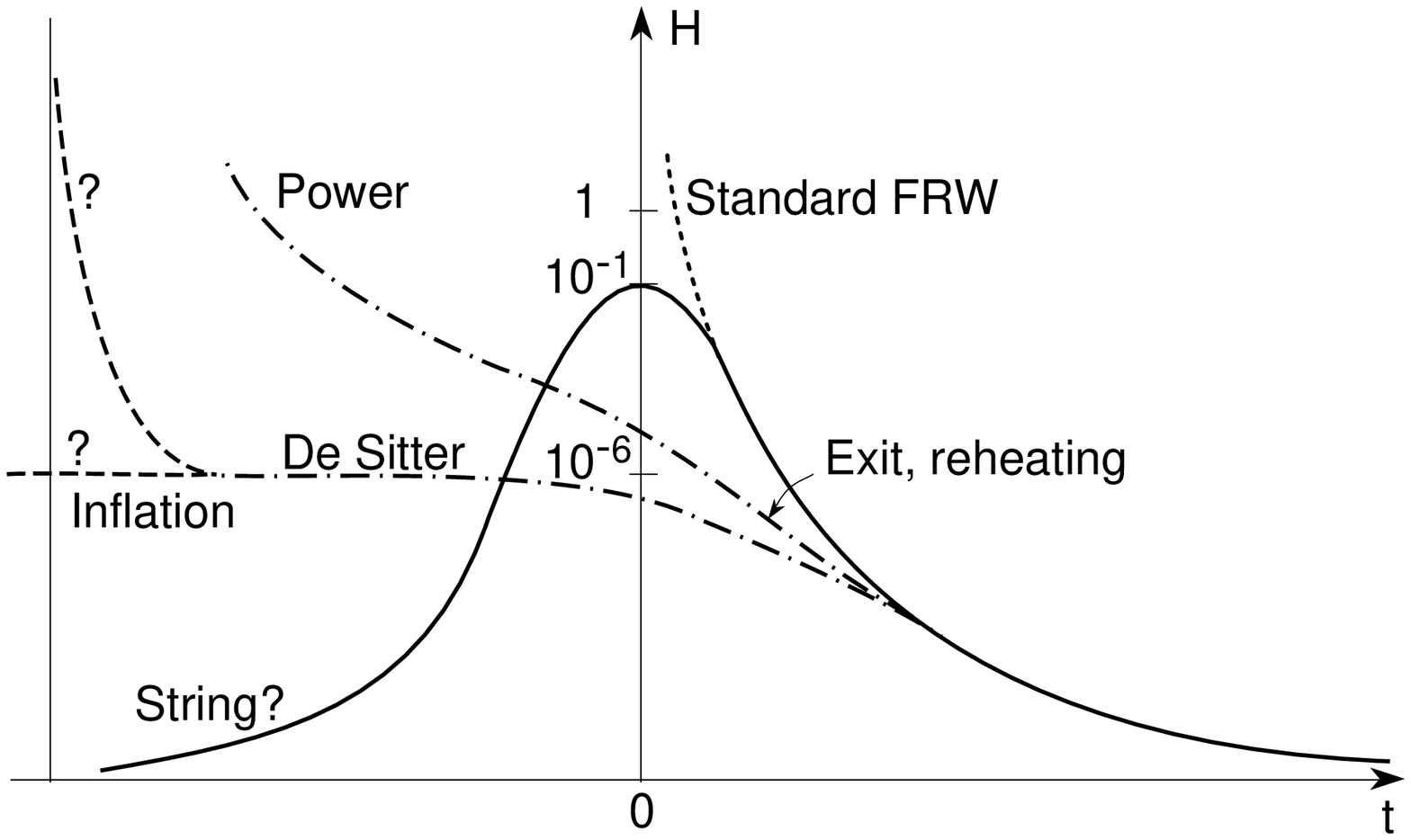,width=10cm}

\noindent
\begin{center}
  {\footnotesize Figure 1}
\end{center}

\hglue 2.0cm \epsfig{figure=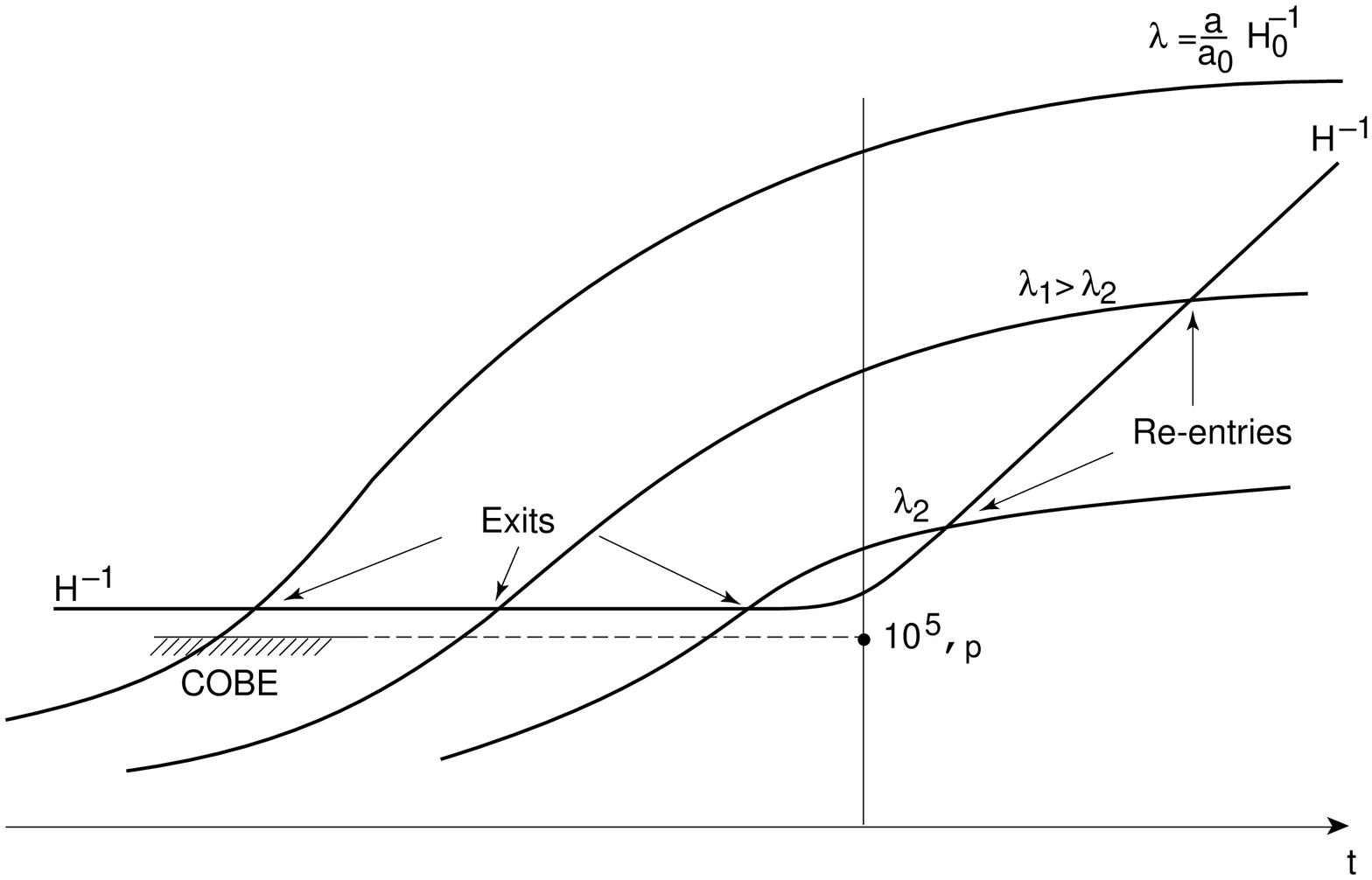,width=10cm}
\begin{center}
  {\footnotesize Figure 2a}\\
\end{center}

\hglue 2.0cm \epsfig{figure=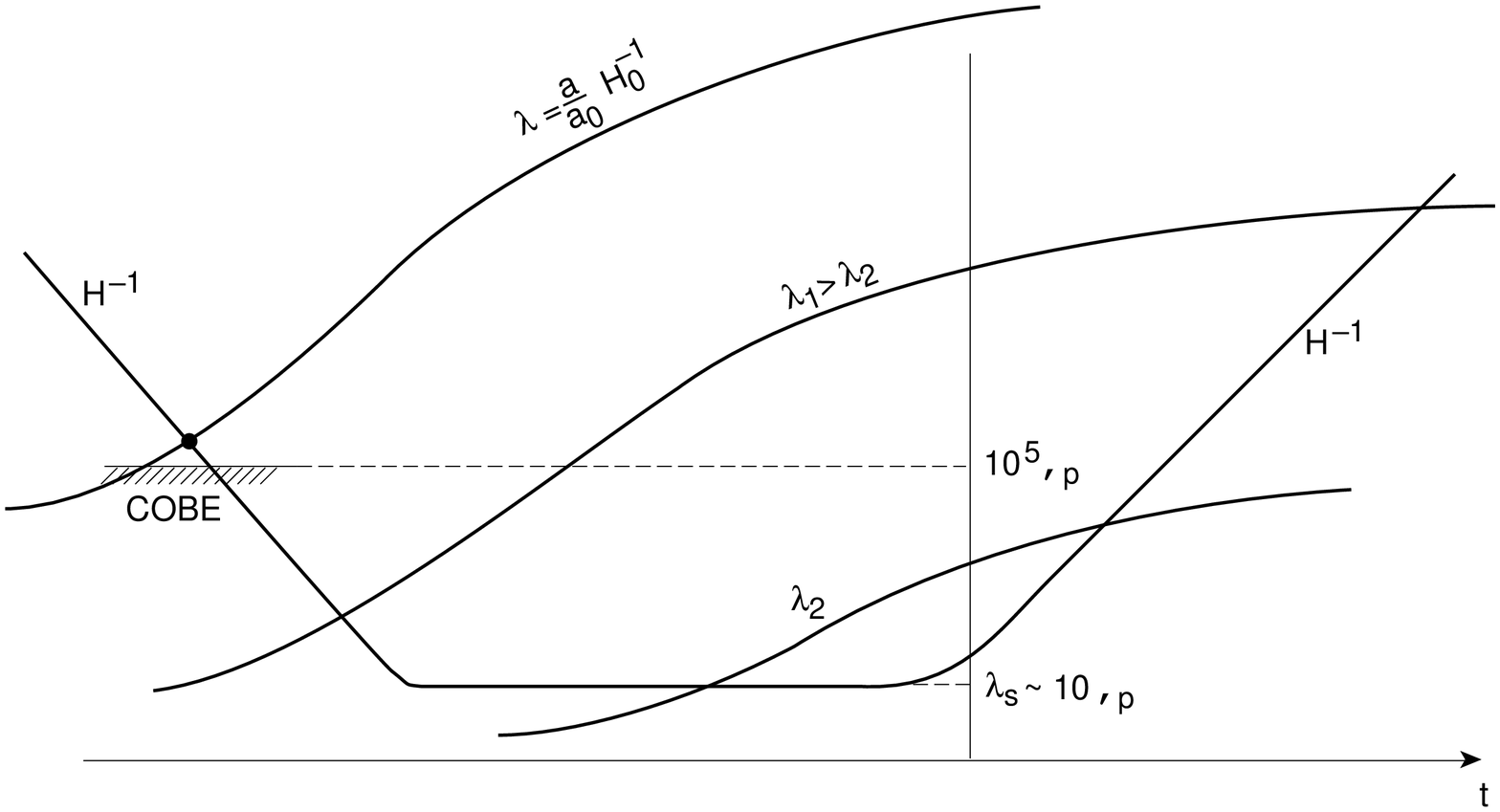,width=10cm}
\begin{center}
  {\footnotesize Figure 2b}
\end{center}

\section {Pre-big bang inflation as a classical instability}

The key to understanding how our apparently innocent initial
conditions can give inflation is in the last attribute we assumed for
the primordial state: the Universe starts deeply inside the
perturbative region, i.e.  at very small coupling. In terms of the
dilaton this means that $\phi$ started very large and negative. This
entitles us to treat the early history of the Universe classically.
Since we have also assumed that it was almost flat, we are also
entitled to use the low-energy approximation to string theory.  All
this can be summarized by saying that we can describe the evolution of
the Universe through the classical field equations of the low-energy
tree-level effective action (\ref{31}); in the simplest case used here
for illustration, this reduces to:
\begin{equation}
\Gamma_{eff} = \frac{1}{2 \lambda^{2}_s} \int d^4x \sqrt{-g}~ e^{-\phi}
 ({\cal R}
+ \partial_\mu \phi \partial^\mu\phi) \; .
\end{equation}
It has been known for some time (see\cite{GV95} for a review) that, if
homogeneity and spatial flatness are assumed, then inflationary
behaviour automatically follows from the stated initial conditions.
Indeed homogeneous, spatially flat solutions fall in {\it four}
categories, of which only {\it one} satisfies the pre-big bang
postulate. The other three exhibit either strong coupling or strong
curvature (or both) in the far past.

However, assuming homogeneity from the start is not very satisfactory.
If we want to solve in a natural way the homogeneity and flatness
problems, we have to start with generic (i.e. not particularly
fine-tuned) initial conditions near the perturbative vacuum. During
the past year this problem has been tackled by several groups, with
somewhat controversial conclusions.  Let me try to explain how I see
the present situation.

Assume that, at some remote time much before the Big Bang, the
Universe was not particularly homogeneous, in the sense that spatial
gradients and time derivatives were both of the same order. Assume
also, in accordance with the PBB postulate, that both kinds of
derivatives were tiny in string units. It can be shown\cite{GV2} that
these initial conditions can lead to a chaotic version of PBB
inflation since, as the system evolves, certain patches develop where
time derivatives slightly dominate over spatial gradients. Provided
this situation is met when the kinetic energy in the dilaton is a
non-negligible fraction of the critical density, dilaton-driven
inflation sets in,\cite{GV2,BMUV1} blowing up the patch and making it
homogeneous, isotropic and spatially flat. The evolution can be
studied by analytical methods (gradient expansion\cite{GE}) since the
approximation of neglecting spatial gradients w.r.t. time derivatives
becomes increasingly accurate within the inflating patch.

The controversial issue is that, in order to have sufficient inflation
in the patch, dilaton-driven inflation has to last sufficiently long.
Its duration is not infinite since it is limited, in the past, by the
conditions I just described and, in the future, by the time at which,
inevitably, curvatures become of string-size and we can no longer
trust the low-energy approximation. Thus, as it was actually noticed
from the very beginning,\cite{GV1} a successful PBB scenario does
require very perturbative initial conditions, so that it takes a long
time (during which the Universe inflates) to reach the BB singularity.
A particular case of this ``fine-tuning" was discussed recently by M.
Turner and E. Weinberg.\cite{TW} They consider a homogeneous, but not
spatially flat Universe and notice that the duration of PBB inflation
is limited in the past by the initial value of the spatial curvature.
This has to be taken very small in string units if sufficient
inflation is to be achieved.

The (almost philosophical) issue is whether this is or is not
fine-tuning.  String theory has a single length parameter,
$\lambda_s$, but, fortunately, it has massless states and low-energy
vacua (such as Minkowsky space-time) whose characteristic scale is
much larger than $\lambda_s$.  Hence I see nothing wrong in starting
the evolution of the Universe in a state of low-energy, small
curvatures, and small coupling. Actually, I find it very amusing that
a classical instability pushes the Universe from low energy
(curvature) and small coupling towards high energy (curvature) and
large coupling.

Another result, which has emerged very recently,\cite{GV2,BMUV1} is
the behaviour of pre-big bang cosmology in the asymptotic past.  If we
evolve the system from the initial conditions I described above
towards the past, we seem to find two possible behaviours.  Either we
reach a singularity at some finite cosmic time in the past, or we flow
smoothly into a trivial space-time.  The first alternative, which
looks generic for positive spatial curvature (e.g.  a $k=+1$ Universe
with $\Omega>1$) has to be excluded since it contradicts our basic
postulate. The second alternative, which looks generic for negative
spatial curvature (e.g. a $k=-1$ Universe with $\Omega<1$), is
perfectly consistent with our philosophy and leads to an interesting
conjecture\cite{BMUV1} for the whole history of time that I will
describe below.  It would be very interesting if pre-big bang
cosmology did predict that the Universe is open, something that
appears to be definitely favoured at present (see e.g.\cite{Science})
by direct measurements of the red-shift-to-distance relation and by
models of large-scale-structure formation.

The complete history at which we arrive can be best drawn on a diagram 
(Fig. 3) reminiscent of (but actually quite different from) a Carter-Penrose
diagram,\cite{CP} which artificially squashes space-time at infinity.
By a simple change of the radial coordinate we can rewrite Eq. (\ref{FRW}) as
\begin{equation}
ds^2 = -dt^2 + a(t)^2~ \left[  dR^2  + r(R)^2 d\Omega^2  \right] \;, 
\label{FRWR}
\end{equation}
and then define the coordinates $x,y$ in  Fig. 3  in terms of $t$ and of the proper distance $aR$ by:
\begin{equation}
\tan (y~\pm~x) = t \pm aR \;.
\label{CP}
\end{equation}
As $ t \rightarrow - \infty$, the Universe approaches an exact vacuum
 of string theory. However, if equal-time hypersurfaces are taken to be those
with a roughly constant energy density, we find that triviality is approached ``a la Milne" i.e. the metric becomes
\begin{equation}
ds^2 = -dt^2 + t^2 \left[ dr^2/(1+r^2) + r^2 d\Omega^2 \right] \; ,
\end{equation}
while the dilaton approaches an arbitrary constant.
For negative time this is just a negative curvature
 FRW Universe, which contracts linearly in time, $a(t) = -t$. For positive
$t$ it is the linearly expanding Universe towards which we are evolving today if $\Omega <1$ (for a discussion of this late-time behaviour see, for instance,
\cite{Zeldovich}). In the Milne Universe the evolution of the scale factor is driven by the negative spatial curvature ($-k~a^{-2} \rightarrow t^{-2}$).
Also shown in Fig. 3 are some time-like geodesics corresponding to a fixed
comoving distance $R$ from the origin.

\hglue 2.0cm \epsfig{figure=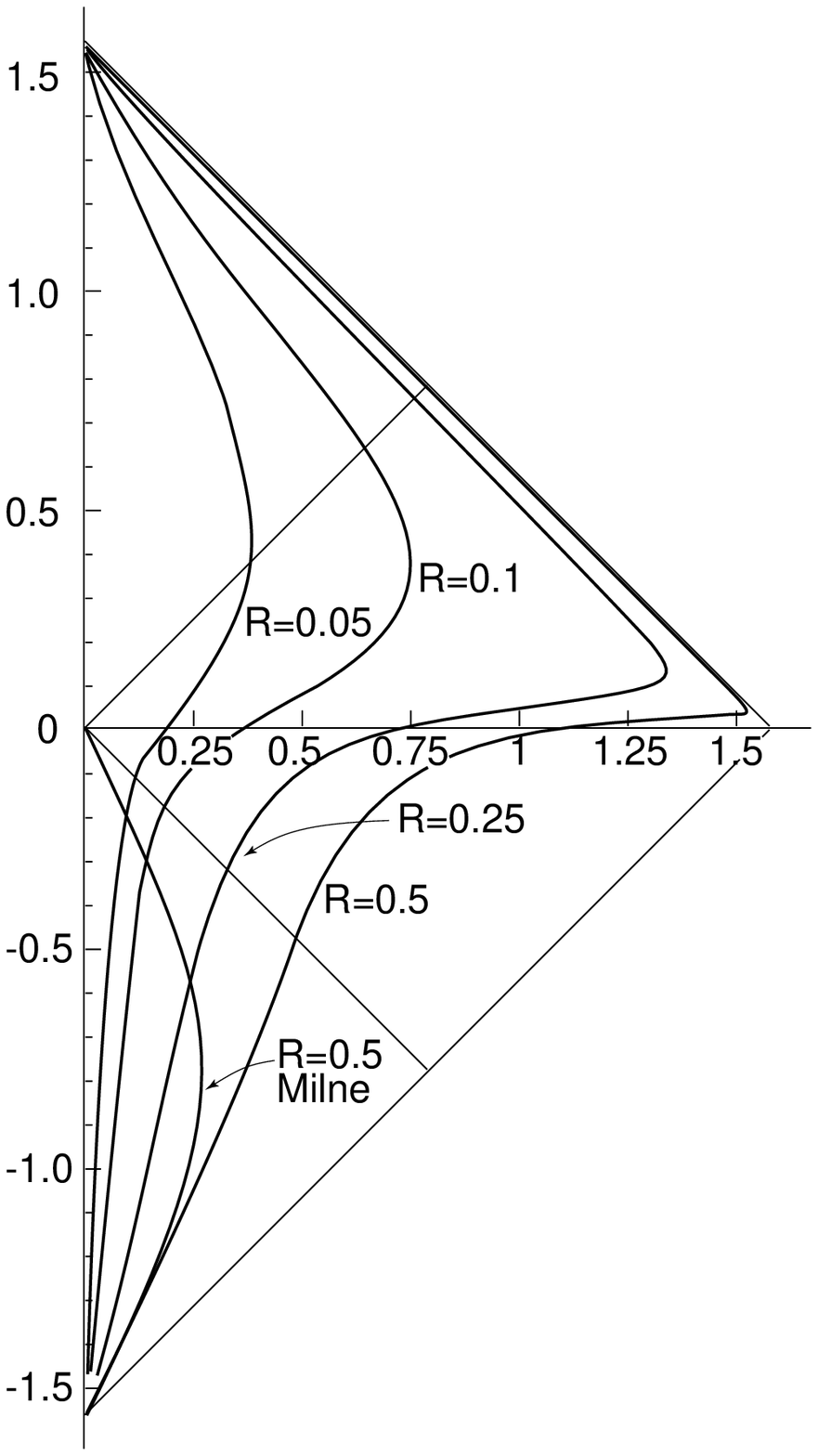,width=8cm}

\noindent
\begin{center}
  {\footnotesize Figure 3}
\end{center}

The generic regular solution thus approaches Milne as $t \rightarrow -
\infty$ but, at any finite large $(-t)$, also contains small dilatonic
(and gravitational-wave) perturbations giving $0< \Omega \ll 1$.  As
$t \rightarrow - \infty$, $\Omega \rightarrow 0$. As time goes
forward, instead, $\Omega$ tends to grow until, at some critical time
$-T_0$, $\Omega$ becomes $O(1)$, in some region of space.  From that
moment on, in that ``lucky" patch, the metric starts to deviate from
Milne (this is shown, in particular, for the $R = 0.5$ geodesic) and
dilaton-driven inflation sets in, pushing $\Omega$ extremely close to
$1$ in that patch.

The rest of the story goes as follows: thanks to the string UV cutoff,
when the curvature becomes $O(\lambda_s^{-2})$, and/or the coupling
becomes $O(1)$, a stringy mechanism prevents reaching the singularity,
and a smooth transition to standard hot big bang FRW cosmology
follows. An interesting quantum mechanism, described in Sec.~5, is
able to provide radiation, temperature and entropy.  However, by then,
the inflated patch is both homogeneous and spatially flat: we have
been able to produce a ``good" big bang!  Post-big bang evolution is
from now on standard, with one qualification.  Although we have
achieved $\Omega =1$ through inflation, we had to start from an open
Universe and thus $\Omega =1-\epsilon ~, ~\epsilon \ll 1$. Inevitably,
FRW evolution will make $\Omega$ deviate more and more from $1$ until,
once more, the Universe will go back to a linearly expanding
Milne-like Universe. It would be wrong to think, however, that the
Universe will just follow the time-reverse of its original life,
firstly because the final coupling is much larger than the initial
one, and, secondly, because entropy has kept increasing all the time:
presumably, in this scenario, the very final stage of the Universe
will consist of an ever increasingly dilute gas of slowly evaporating
black holes...

\section{ Quantum Mechanical Heating of the Universe and Observable PBB relics}

Since there are already several review lectures on this subject
(e.g.\cite{GV95,MG95}), I will limit myself to mention the most recent
developments simply after recalling the basic physical mechanism
underlying particle production in cosmology.\cite{quantum} A
cosmological (i.e. time-dependent) background coupled to a given type
of (small) inhomogeneous perturbation $\Psi$ enters the effective
low-energy action in the form:
\begin{equation}
I ={\small\frac{1}{2}} \int d\eta\ d^3x\ S(\eta) \left[ \Psi
^{\prime 2}- (\nabla \Psi )^2\right].
\label{spertact}
\end{equation}
Here $\eta$ is the conformal time coordinate, and a prime denotes
$\partial/\partial\eta$. The function $S(\eta)$ (sometimes called the
``pump" field) is, for any given $\Psi$, a given function of the scale
factor, $a(\eta)$, and of other scalar fields (four-dimensional
dilaton $\phi(\eta)$, moduli $b_i(\eta)$, etc.) which may appear
non-trivially in the background.

While it is clear that a constant pump field $S$ can be reabsorbed in
a rescaling of $\Psi$, and is thus ineffective, a time-dependent $S$
couples non-trivially to the fluctuation and leads to the production of
pairs of quanta (with opposite momenta).  Looking back at Eq.
(\ref{31}), one can easily determine the pump fields for each one of
the most interesting perturbations.  The result is:
\begin{eqnarray}
\rm{Gravity~waves,~dilaton} &:&   S = a^2 e^{-\phi} \nonumber \\
\rm{Heterotic~gauge~bosons} &:&  S =  e^{-\phi} \nonumber \\
\rm{Kalb-Ramond,~axions} &:&   S = a^{-2} e^{-\phi}
\end{eqnarray}

A distinctive property of string cosmology is that the dilaton $\phi$
appears in some very specific way in the pump fields. The consequences
of this are very interesting:
\begin{itemize}
\item For gravitational waves and dilatons the effect of $\phi$ is to
  slow down the behaviour of $a$ (remember that both $a$ and $\phi$
  grow in the pre-big bang phase).  This is the reason why those
  spectra are quite steep\cite{BGGV} and give small contributions at
  large scales.
\item For (heterotic) gauge bosons there is no amplification of vacuum
  fluctuations in standard cosmology, while, in string cosmology, all
  the ``work" is done by the dilaton.  In the pre-big bang scenario,
  the coupling must grow by as large a factor as the one by which the
  Universe has inflated. This implies a very large amplification of
  the primordial quantum fluctuation,\cite{GGV} possibly explaining
  the long-sought origin of seeds for the galactic magnetic fields.
\item Finally, for Kalb-Ramond fields and axions, $a$ and $\phi$ work
  in the same direction and spectra can be large even at large
  scales.\cite{Copeland} Note, incidentally, that the power of $a$ in
  $S$ is determined by the rank of the corresponding tensor.  It is
  well known, however, that the Kalb-Ramond field can be reduced to a
  (pseudo)scalar field, the axion, through a duality transformation.
  This turns out to change $S$ into $S^{-1}$, i.e.  the pump field for
  the axion is actually $a^{2} e^{\phi}$.  An interesting duality of
  cosmological perturbations, reminiscent of electric-magnetic (or
  strong-weak) duality, can be argued\cite{duality} to guarantee the
  equivalence of the Kalb-Ramond and axion spectra.
\item Many other fluctuations, which arise in generic
  compactifications of superstrings, have also been studied and lead
  to interesting spectra. For lack of time, I will refer to the
  existing literature.\cite{BH,BMUV2}
\end{itemize}

The possible flatness of axionic spectra in pre-big bang cosmology
leads to hopes that, in such a scenario, there is a natural way to
generate an interesting spectrum of large-scale fluctuations, one of
the much advertised properties of the standard inflationary scenario.
Work is still in progress to establish whether this hope is indeed
realized.

Before closing this section, I wish to recall how one sees the very
origin of the hot big bang in this scenario. One can easily estimate
the total energy stored in the quantum fluctuations which were
amplified by the pre-big bang backgrounds. The result is, roughly,
\begin{equation}
\rho_{quantum} \sim N_{eff} ~ H^4_{max} \; ,
\end{equation}
where $N_{eff}$ is the effective number of species that are amplified
and $H_{max}$ is the maximal curvature scale reached around $t=0$
(this formula has to be modified in case some spectra show negative
slopes). We have already argued that $H_{max} \sim \lambda_s^{-1}$ and
we know that, in heterotic string theory, $N_{eff}$ is in the
hundreds. Yet this rather huge energy density is very far from
critical as long as the dilaton is still in the weak-coupling region,
justifying our neglect of back-reaction effects. It is very tempting
to assume that, precisely when the dilaton reaches a value such that
$\rho_{quantum}$ is critical, the Universe will enter the
radiation-dominated phase. This too is, at present, the object of
active investigation.

\newpage
\section{Conclusion}

Pre-big bang cosmology appears to have survived its first 6-7 years of
life. Interest in (criticism of) it is clearly growing. It is perhaps
time to make a balance sheet.

Conceptual (technical?) and phenomenological problems include:
\begin{itemize}
\item Graceful exit from dilaton-driven inflation to FRW cosmology is
  not fully understood, in spite of recent progress.\cite{exit}
  Possibly, new ideas borrowed from M-theory and D-branes could help
  in this respect.\cite{branes}
\item A scale-invariant spectrum of large-scale perturbations does not
  look automatic, although, for the first time, thanks to the flat
  axion spectra, it does not look impossible either.
\end{itemize}

Attractive features include:
\begin{itemize}
\item No need to ``invent" an inflaton, or to fine-tune potentials.
\item Inflation is ``natural" thanks to the duality symmetries of
  string cosmology.
\item The initial conditions problem is decoupled from the singularity
  problem: a solution to the former is already shaping up and looks
  exciting.
\item A classical gravitational instability finds a welcome use in
  providing inflation; a quantum instability (pair creation) is able
  to heat up an initially cold Universe and generate a standard hot
  big bang with the additional features of homogeneity, flatness and
  isotropy.
\item Last but not least: one is dealing with a highly constrained,
  predictive scheme which can be tested/falsified by low-energy
  experiments thanks to the fact that a huge red-shift has brought the
  scale of Planckian physics down to that of human beings:
\begin{equation}
(l_P/H_0)^{1/2} \sim 1~\rm{mm} \nonumber
\end{equation}

\end{itemize}

\noindent
{\bf Note Added}

Since I gave this lecture, two relevant papers have appeared:
\begin{itemize}
\item[1)] N. Kaloper, A. Linde and R. Bousso (hep-th/9801073) have
  added further points to the criticism of
the PBB scenario expressed in Ref.~16.
\item[2)] A numerical study of the spherically symmetric case by J.
  Maharana, E. Onofri and G. Veneziano (gr-qc/9802001) appears to
  support the idea discussed in Section~4 that PBB behaviour emerges
  generically from initial conditions sufficiently close to Milne's
  trivial vacuum.
\end{itemize}

These two papers confirm that much more work is still needed to
clarify all the relevant issues raised by the new cosmological setup
I~discussed here.

\end{document}